\documentclass[aps,superscriptaddress, twocolumn,showpacs, longbibliography]{revtex4-1}
\pdfoutput=1
\usepackage{color}
\usepackage{amsmath, amsthm, amsfonts}    
\usepackage{amssymb}
\usepackage{mathtools}
\usepackage{framed} 
\usepackage{mathptmx} 
\usepackage[table]{xcolor}
\usepackage{dsfont}
\usepackage{enumitem} 
\usepackage[]{mdframed}
\usepackage{appendix}
\usepackage{braket}
\usepackage[skins,theorems]{tcolorbox}
\tcbset{highlight math style={enhanced,
  colframe=red,colback=white,arc=0pt,boxrule=1pt}}
\usepackage{cancel}
\usepackage{empheq}
\usepackage{lipsum}
\usepackage{graphicx}
\usepackage{tikz}
\usepackage{mdframed}
\usepackage{bigints}
\usepackage{makecell}
\usepackage[unicode=true,pdfusetitle,
 bookmarks=true,bookmarksnumbered=false,bookmarksopen=false,
 breaklinks=true,pdfborder={0 0 0},backref=false,colorlinks=true,citecolor=blue]{hyperref}
\usepackage{graphicx}   
\usepackage[caption=false]{subfig}       
\usepackage{bm}            
\usepackage[normalem]{ulem}  

\newlength\imagewidth
\newlength\imagescale
\def\be{\begin{eqnarray}}
\def\ee{\end{eqnarray}}
\def\r{{\bf r}}

\def\E{{\bf E}}
\usepackage{tabstackengine}
\setstackEOL{\cr}

\def\S{{\bf S}}

\AtBeginEnvironment{pmatrix}{\setlength{\arraycolsep}{1pt}}

\DeclareMathOperator{\atantwo}{atan2}

\DeclareUnicodeCharacter{2212}{-}

\definecolor{JOT-color}{named}{blue}
\definecolor{CSF-color}{named}{orange}

\begin{document}

\title{Solving Maxwell's Equations Using  Polarimetry Alone}

\author{Jorge Olmos-Trigo}
\email{jolmostrigo@gmail.com}
\affiliation{Departamento de Física, Universidad de La Laguna, Apdo. 456. E-38200, San Cristóbal de La Laguna, Santa Cruz de Tenerife, Spain.}

\begin{abstract}
Maxwell's equations are solved when the amplitude and phase of the electromagnetic field are determined at all points in space. Generally, the Stokes parameters can only capture the amplitude and polarization state of the electromagnetic field in the radiation (far) zone. 
Therefore, the measurement of the Stokes parameters is, in general, insufficient to solve Maxwell's equations. In this Letter, we solve Maxwell's equations for a set of objects widely used in Nanophotonics using the Stokes parameters alone.   Our method for solving Maxwell's equations endows the Stokes parameters an even more fundamental role in the electromagnetic scattering theory. 
\end{abstract}

\maketitle

{\emph{Introduction.}}---The determination of the amplitude and phase of the electromagnetic field at all points in space solves Maxwell's equations~\cite{maxwell1865viii}.
Currently, electromagnetic software packages can provide the numerical solution to Maxwell's equations~\cite{multiphysics1998introduction}.
However, solving Maxwell's equations in the optical laboratory is nearly infeasible. One needs to measure the components of the scattered field at all points of the radiation zone. On top of that,  the internal field induced in the excited object is experimentally inaccessible.

In stark contrast, the Stokes parameters can be readily measured using a photodiode and  waveplates~\cite{stokes1851composition, bohren2008absorption}. The Stokes parameters 
can capture the amplitude and polarization state of the electromagnetic field in the radiation (far) zone.
Following the notation of Ref.~\cite{bohren2008absorption}, we can write the Stokes parameters as
\begin{alignat}{2} 
s_0 &= |E_\theta|^2 + |E_\varphi|^2, & \qquad
\label{s_01}
s_1 &= |E_\theta|^2 - |E_\varphi|^2, \\
s_2 &= -2\Re \{E_\theta E^*_\varphi \}, & \qquad
s_3 &= 2\Im \{E_\theta E^*_\varphi \}.
\label{s_23}
\end{alignat}
As Eqs.~\eqref{s_01}-\eqref{s_23} show, the Stokes parameters depend on the transversal components of the scattered electromagnetic field evaluated in the radiation (far) zone, i.e., $E_\theta$ and $E_\varphi$~\footnote{Note that the longitudinal component of the scattered electromagnetic fields identically vanishes in far-field, namely, $E_r = 0$}. If the amplitude and phase of  $E_\theta$ and $E_\varphi$ are measured, the Stokes parameters can be calculated using Eqs.~\eqref{s_01}-\eqref{s_23}~\cite{crichton2000measurable, marston2018humblet}. The converse is not generally true. 
To demonstrate this fact, we now write $E_\theta = |E_\theta| e^{ i \xi_\theta}$ and $E_\varphi = |E_\varphi| e^{ i \xi_\varphi}$, where $\xi_\theta$ and $\xi_\varphi$ are real-valued phases. Taking into account this notation, we can obtain the following relations from Eqs.~\eqref{s_01}-\eqref{s_23}
\begin{align} \label{good}
|E_\theta|^2 = \frac{s_0 + s_1}{2}, && |E_\varphi|^2 = \frac{s_0 - s_1}{2},
\end{align}
\begin{equation} \label{ind}
\tan (\xi_{\theta} -\xi_{\varphi}) = {s_3}/{s_2}.   
\end{equation}
On the one hand, Eq.~\eqref{good} reveals that the measurement of $s_0$ and $s_1$ grants access to the amplitudes $|E_\theta|$ and $|E_\varphi|$. On the other hand, Eq.~\eqref{ind} shows that measuring $s_2$ and $s_3$ provides the phase difference $\xi_{\theta} -\xi_{\varphi}$ but falls short of providing the individual phases  $\xi_{\theta}$ and $\xi_{\varphi}$. As we mentioned earlier, the determination of the phase of the electromagnetic field is necessary to solve Maxwell's equations. Therefore, one could conclude that measuring the Stokes parameters is insufficient to solve Maxwell's equations. 

In this Letter, we demonstrate that a measurement of the Stokes parameters at a single scattering angle is sufficient to solve Maxwell's equations for a set of objects. These objects share the following features: they are lossless, axially-symmetric and their optical response is well-described by a single multipole order. Notably, several works have tackled such objects in different branches of Nanophotonics. Examples include optically resonant nanoantennas~\cite{kuznetsov2012magnetic, evlyukhin2012demonstration, sheikholeslami2013metafluid, zywietz2014laser, shima2023gallium}, Kerker conditions~\cite{geffrin2012magnetic, fu2013directional, staude2013tailoring, person2013demonstration}, surface-enhanced Raman scattering~\cite{dmitriev2016resonant}, surface-enhanced optical chirality~\cite{negoro2023helicity, olmosfar}, among many others~\cite{chaabani2019large, ishii2017resonant, sugimoto2017colloidal, sugimoto2020mie}. Note that Refs.~\cite{kuznetsov2012magnetic, evlyukhin2012demonstration, sheikholeslami2013metafluid, zywietz2014laser, shima2023gallium, geffrin2012magnetic, fu2013directional, staude2013tailoring, person2013demonstration,
dmitriev2016resonant, negoro2023helicity, olmosfar, ishii2017resonant,chaabani2019large,ishii2017resonant, sugimoto2017colloidal, sugimoto2020mie} are {experimental studies} widely recognized by the Nanophotonics community.

The key to our procedure lies in linking the measurement of the Stokes parameters in the radiation (far) zone with the electric and magnetic scattering coefficients of the multipolar expansion of the scattered field. As we show, the determination of these scattering coefficients solves Maxwell's equations at all points of the radiation zone, ranging from far-to-near field. 
Additionally, in the case of spherical objects,  we solve Maxwell's equations at all points in space (also inside the object). 



{\emph{The multipolar expansion of the fields.}}---We now consider the scattered ${\E_{\rm{sca}}(k \r)}$ and internal ${\E_{\rm{int}}(k \r)}$ electromagnetic fields produced by an arbitrary object. In the usual basis of electric and magnetic multipoles~\cite{jackson1999electrodynamics}, we can write the scattered and internal electromagnetic fields  as
\begin{equation} \label{E_sca}
{\E_{\rm{sca}}(k \r)}= E_0 \sum_{\ell = 1}^{\infty} \sum_{m = -\ell }^{\ell}  a_{\ell m}\boldsymbol{N}^{h}_{\ell m}(k \r) +  b_{\ell m}\boldsymbol{M}^{h}_{\ell m}(k \r),
\end{equation}
\begin{equation} \label{E_int}
{\E_{\rm{int}}(k_i \r)}= E_0 \sum_{\ell = 1}^{\infty} \sum_{m = -\ell }^{\ell}  d_{\ell m}\boldsymbol{N}^{j}_{\ell m}(k_i \r) +  c_{\ell m}\boldsymbol{M}^{j}_{\ell m}(k_i \r).
\end{equation}
Here $\boldsymbol{M}^{h}_{\ell m}(k \r) = h^{(1)}_\ell \boldsymbol{X}_{\ell m}(\theta, \varphi)$ and $\boldsymbol{M}^{j}_{\ell m}(k \r) = j_\ell \boldsymbol{X}_{\ell m}(\theta, \varphi)$, $h^{(1)}_\ell$ and $j_\ell$ are spherical Hankel and Bessel functions of the first kind, respectively, $\boldsymbol{X}_{\ell m}(\theta, \varphi)$ represents the usual vector spherical harmonics~\cite{jackson1999electrodynamics}, $k \boldsymbol{N}^{s}_{\ell m}(k \r) = i \mathbf{\nabla} \times \boldsymbol{M}^{s}_{\ell m}(k \r)$, where $s = \{j, h \}$, and
$\ell$ and $m$ denote the multipolar order and the total angular momentum, respectively. 
Moreover $\r = \{r, \theta, \varphi \}$ is the observational point, $k$ is the radiation wavenumber, $k_i = {\rm{m}} k$, $\rm{m}$ being the refractive index of the object, and $E_0$ is the amplitude of the incident wavefield.
Importantly, $a_{\ell m}$ and  $b_{\ell m}$ denote the electric and magnetic scattering coefficients, respectively, and $d_{\ell m}$ and $c_{\ell m}$ are the internal electric and magnetic coefficients, respectively.

Equations~\eqref{E_sca}-\eqref{E_int} show that the determination of the set $\{a_{\ell m}, b_{\ell m}, d_{\ell m}, c_{\ell m} \}$ grants access to the amplitude and phase of both the scattered and internal electromagnetic fields at all points of space. 
However, capturing the set $\{a_{\ell m}, b_{\ell m}, d_{\ell m}, c_{\ell m} \}$  is exceptionally demanding due to the need to measure the components of the scattered and internal electromagnetic fields in all directions~\cite{jackson1999electrodynamics}.  In fact and to the best of our knowledge, none of the magnitudes of the set has been experimentally measured.

{\emph{The Stokes parameters measurement and its limitations.}}---As previously mentioned, the Stokes vector $\S = \{s_0, s_1, s_2, s_3 \}$ can be measured in the radiation (far) zone with conventional optical components such as a photodiode and waveplates. Hereafter, we consider objects well-described by fixed values of $m$ and $\ell$. In other words, we deal with axially symmetric objects whose optical response is described by a single multipolar order. We recall that such objects have been widely studied in Nanophotonics~\cite{kuznetsov2012magnetic, evlyukhin2012demonstration, sheikholeslami2013metafluid, zywietz2014laser, shima2023gallium, geffrin2012magnetic, fu2013directional, staude2013tailoring, person2013demonstration,
dmitriev2016resonant, negoro2023helicity, olmosfar, ishii2017resonant,chaabani2019large,ishii2017resonant, sugimoto2017colloidal, sugimoto2020mie}.
In this setting ($\ell m$ is fixed), let us insert the far-field limit ($kr \rightarrow \infty$) of Eq.~\eqref{E_sca} into Eqs.~\eqref{s_01}-\eqref{s_23}. After some algebra (see Supporting material S1), we get~\cite{olmos2023stokes}
%

\begin{equation} \label{compact}
 \begin{pmatrix}
    |a_{\ell m}|^2\\
    |b_{\ell m}|^2 \\ \Re \{a_{\ell m} b^*_{\ell m} \}\\
    \Im \{a_{\ell m} b^*_{\ell m} \}
    \end{pmatrix}=  
U_{\ell m}
 \begin{pmatrix}
    {s}_0\\
    {s}_1\\ {s}_2\\
    {s}_3
    \end{pmatrix}.
\end{equation}
Equation~\eqref{compact} shows that all the quadratic combinations of  $\{a_{\ell m}, b_{\ell m}\}$ can be attained from a single Stokes vector measurement in the far-field. As proved in Ref.~\cite{olmos2023stokes}, all that one needs to do is compute the 4x4 matrix $U_{\ell m}$ and apply it to the Stokes measurement.
However, we anticipate that even if the object is well-described by fixed values of $m$ and $\ell$, the phases of $a_{\ell m}$ and $b_{\ell m}$ cannot be attained using  Eq.~\eqref{compact}. 
To prove it, we now write $a_{\ell m} = |a_{\ell m}|e^{ i \phi_a}$ and $b_{\ell m} = |b_{\ell m}|e^{ i \phi_b}$, where $\phi_a$ and $\phi_b$ are real-valued phases. In this setting, the last two rows of the left side of Eq.~\eqref{compact} can be manipulated to yield 
\begin{equation} \label{und}
 \tan (\phi_a  -\phi_b) = \frac{\Im \{a_{\ell m} b^*_{\ell m} \}}{\Re \{a_{\ell m} b^*_{\ell m} \}}.
\end{equation}
Equation~\eqref{und} provides access to the phase difference $\phi_a - \phi_b$ but not to the individual phases $\phi_a$ and $\phi_b$.  Consequently, without knowledge of these individual phases, determining the scattering coefficients $a_{\ell m}$ and $b_{\ell m}$ becomes impossible. Due to this infeasibility, we cannot capture the scattered  field $\E_{\rm{sca}}(k \r)$  using Eq.~\eqref{compact}, and the solution to Maxwell's equations is not attained. 

\begin{figure*}[t!]
    \centering
    \includegraphics[width=\textwidth]{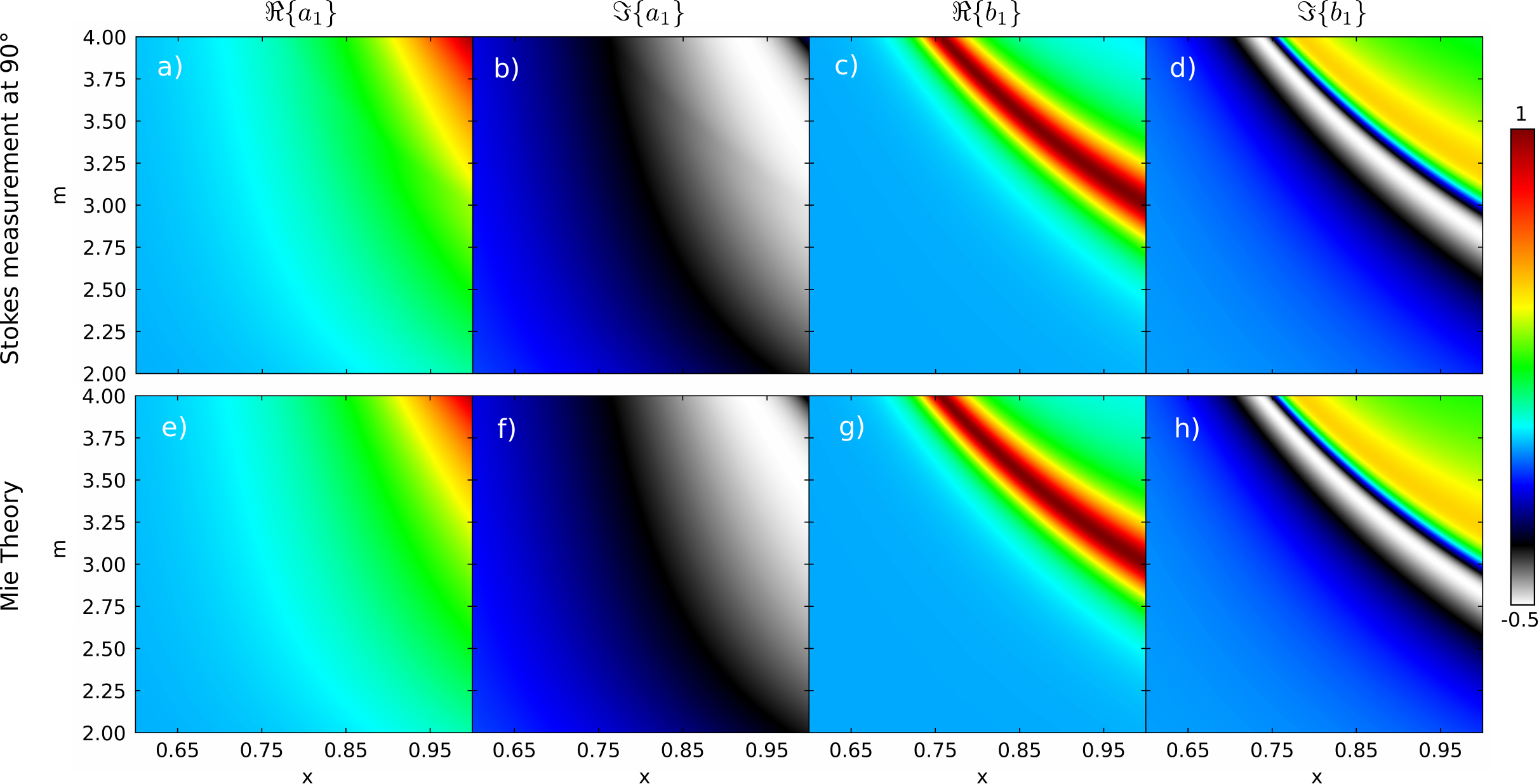}
    \caption{Real and imaginary parts of the dipolar electric and magnetic Mie coefficients obtained from both a Stokes measurement at $\theta = 90^\circ$ (see Figs. a-d) and using exact Mie theory (see Figs. e-h). The excitation wavefield is a circularly polarized plane wave in both cases. The real and imaginary parts of the  Mie coefficients are depicted vs the refractive index contrast $\rm{m}$ and the optical size $x = ka = (2 \pi a) / \lambda, $, $\lambda$ and a being the radiation wavelength and the radius of the spherical nanoparticle. The intense red colors indicate Mie resonances.}
    \label{Full}
\end{figure*}

\begin{figure*}[t!]
    \centering
    \includegraphics[width=1\textwidth]{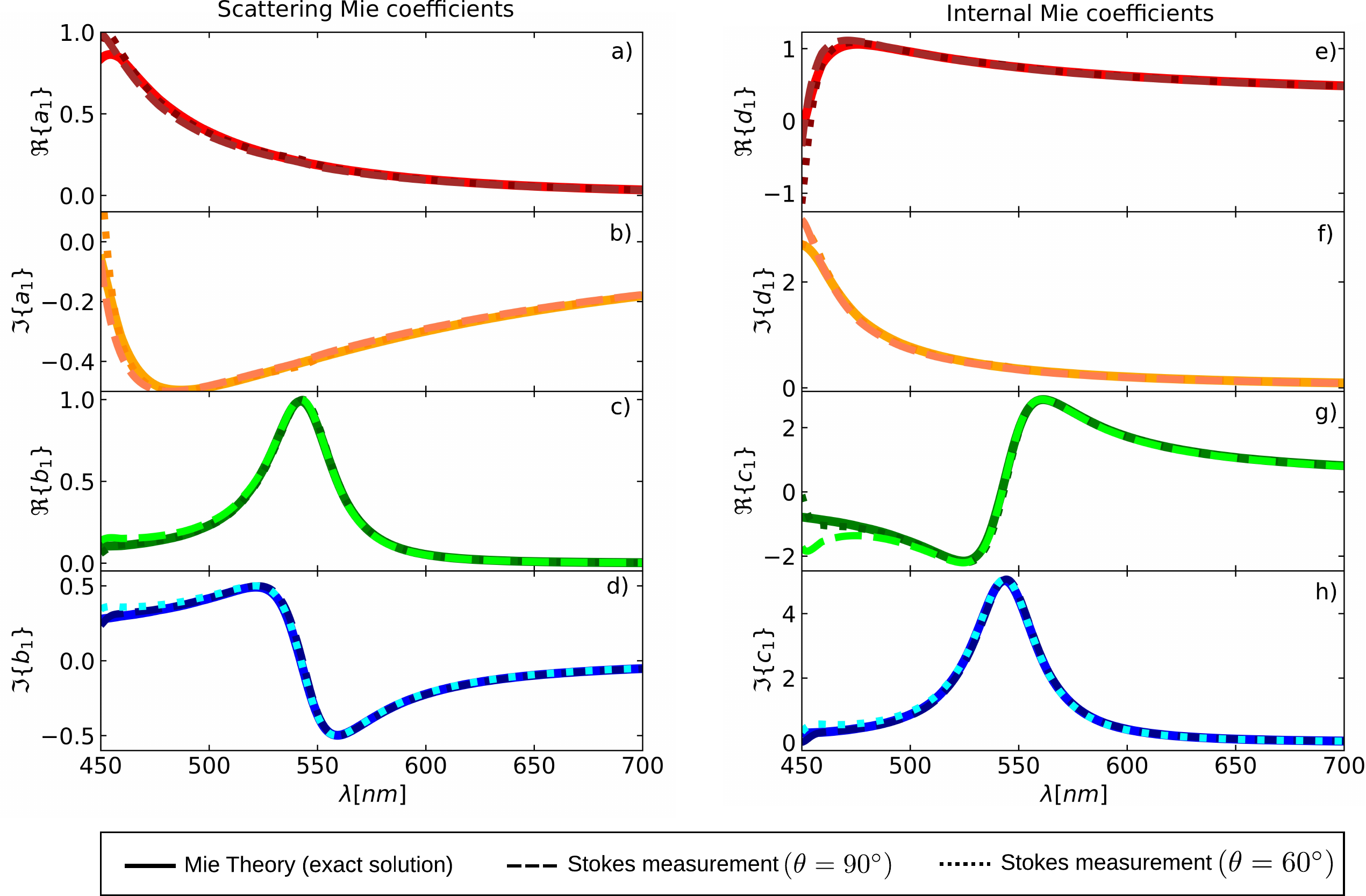}
    \caption{Real and imaginary parts of the scattering (see Figs. a-d) and internal (see Figs. e-h)  Mie coefficients of a GaP spherical object with radii $a = 75$ nm obtained from both a Stokes measurement at $\theta =  90^\circ$ (dashed) and $\theta =  60^\circ$ (dotted). The excitation wavefield is a circularly polarized plane wave in all cases. The scattering and internal  Mie coefficients are depicted vs the incident wavelength $\lambda$. }
    \label{GaP}
\end{figure*}

In the forthcoming, we show that the phase-indetermination of Eq.~\eqref{und} is resolved if the light-scattering system is lossless.

{\emph{Unveiling the scattered field at all points of the radiation zone.}}---We now consider the extinction and scattering cross-sections, denoted by 
$\sigma_{\rm{ext}} = \sigma^{\rm{e}}_{\rm{ext}} +  \sigma^{\rm{m}}_{\rm{ext}}$ and $\sigma_{\rm{sca}} = \sigma^{\rm{e}}_{\rm{sca}} + \sigma^{\rm{m}}_{\rm{sca}}$, respectively. Note that here $\rm{e}$ and $\rm{m}$ denote the electric and magnetic contributions, respectively. 
The extinction and scattering cross-sections can be written as~\cite{mishchenko2002scattering}
\begin{align} \label{sec_e}
k^2 \sigma^{\rm{e}}_{\rm{ext}} &= - \Re \{{g^{\rm{e}}_{\ell m}} a^*_{\ell m}\}, & k^2 \sigma^{\rm{m}}_{\rm{ext}} &= - \Re \{{g^{\rm{m}}_{\ell m}} b^*_{\ell m}\}, 
\\ \label{sec_m}
k^2 \sigma^{\rm{e}}_{\rm{sca}} &= |a_{\ell m}|^2, & k^2 \sigma^{\rm{m}}_{\rm{sca}} &= |b_{\ell m}|^2. 
\end{align}
Here $g^{\rm{e}}_{\ell m}$ and $g^{\rm{m}}_{\ell m}$ denote the beam-shape coefficients of the incident wavefield~\cite{zambrana2012excitation}.
Objects without optical losses satisfy $\sigma_{\rm{ext}} = \sigma_{\rm{sca}}$. According to Eqs.~\eqref{sec_e}-\eqref{sec_m}, lossless objects fulfill
\begin{align} \label{phase_amp}
|a_{\ell m}|^2 =  -\Re \{a^*_{\ell m} g^{\rm{e}}_{\ell m} \}, && |b_{\ell m}|^2 =  -\Re \{b^*_{\ell m} g^{\rm{m}}_{\ell m} \}.
\end{align}
Equation~\eqref{phase_amp} shows that the amplitude of the electric (and magnetic) scattering coefficient is a function of the electric (and magnetic) phase~\cite{hulst1957light, olmos2020unveiling, olmos2022helicity}. 
That noted, by expanding Eq.~\eqref{phase_amp} and manipulating Eq.~\eqref{und}, we arrive to
\begin{align}
& \Re \{g^{\rm{e}}_{\ell m} \}\cos \phi_a + \Im \{g^{\rm{e}}_{\ell m} \}\sin \phi_a = -|a_{\ell m}|, \label{eq:equation1} \\
& \Re \{g^{\rm{m}}_{\ell m} \}\cos \phi_b + \Im \{g^{\rm{m}}_{\ell m} \}\sin \phi_b = -|b_{\ell m}|, \label{eq:equation2} \\
& (\phi_a - \phi_b) = \atantwo \left[ \frac{\Im \{a_{\ell m} b^*_{\ell m} \}}{\Re \{a_{\ell m} b^*_{\ell m} \}} \right]. \label{eq:equation3}
\end{align} 
We now reach notable results. The system of Eqs.~\eqref{eq:equation1}-\eqref{eq:equation3} can be unambiguously solved yielding $\phi_a$ and $\phi_b$ (in the correct quadrant). Note that the right-side of Eqs.~\eqref{eq:equation1}-\eqref{eq:equation3} can be obtained using Eq.~\eqref{compact}. Moreover,  the beam-shape coefficients $\{g^{\rm{e}}_{\ell m}, g^{\rm{m}}_{\ell m}   \}$ are usually known quantities since the incident electromagnetic field can be controlled.
Now, capturing $\phi_a$ and $\phi_b$ along with the amplitudes $|a_{\ell m}|$ and $|b_{\ell m}|$   allows us to determine the electric and magnetic  scattering coefficients $a_{\ell m}$ and $b_{\ell m}$.
As Eq.~\eqref{E_sca} shows, the determination of $a_{\ell m}$ and $b_{\ell m}$ grants access to all the components of the scattered  field evaluated at all points of the radiation zone, ranging from far-to-near-field. In simple words, the determination of $a_{\ell m}$ and $b_{\ell m}$ solves Maxwell's equations in the radiation zone.

At this point, let us highlight the main features of our method for solving Maxwell's equations in the radiation zone:
\begin{itemize}
    \item {Simplicity in the measurement}: Our method relies on a measurement of the Stokes parameters at a single angle. From an experimental standpoint, we must avoid any propagation direction where the incident wavefield has a component. Otherwise, the total field measured in the optical laboratory will be the sum of the scattered and incident wavefields, thus invalidating our method. 

    \item {Generality of the material and shape of the object}: Our method works for axially symmetrical objects such as disks, pillars, spheres, and spheroids. Note that these objects can be composed of a single material or not (coated objects). The only requirement is that the absorption cross-section of such objects must be zero.   

 \item {Wide range of illumination conditions}: Our method can accommodate plane waves, Gaussian beams, or even vortex wavefields~\cite{zambrana2012excitation, zambrana2013dual}. The only requirement is that the total angular momentum $m$ of the light-scattering system must be well-defined.
\end{itemize}

{\emph{Capturing the scattering Mie coefficients.}}--- At this point, let us illustrate the relevance of Eqs.~\eqref{eq:equation1}-\eqref{eq:equation3} with one of the most canonical examples used in Nanophotonics: all-dielectric spherical nanoparticles excited by a plane wave. For details on the beam-shape coefficients of a plane wave, check Supplementary Material S2. Additionally, check Supplementary Material S3 to learn how Eqs.~\eqref{eq:equation1}-\eqref{eq:equation3} simplify for spherical and lossless particles.



First, we illustrate the accuracy of our method to capture $a_1$ and $b_1$. In Fig.~\ref{Full}, we depict $\Re \{a_1\}$, $\Im \{a_1\}$, $\Re \{b_1\}$, and $\Im \{b_1\}$ calculated from a Stokes measurement at the scattering angle $\theta = 90^\circ$ and utilizing Mie theory (exact solution)~\footnote{Note that $\varphi$ does not play a role due to the symmetries of the light-scattering system. Additionally, notice that other scattering angles $\theta$ could have been selected.}. 
The calculation of the Mie coefficients obtained from the Stokes measurement shows an excellent agreement with the exact solution in the broadband interval of refractive index contrasts $2<\rm{m}<4$ and optical sizes  $0.6$ $<\rm{x}< 1$. Here $\rm{x} = ka = 2 \pi a / \lambda$,  $\lambda$ and a being the radiation wavelength and the radius of the object, respectively. Note that for $0<x<0.6$, our approach, summarized in Eqs.~\eqref{eq:equation1}-\eqref{eq:equation3}, works as it holds for objects described by an electric and/or magnetic response. 

Let us stress that the scattering Mie coefficients $a_\ell$ and $b_\ell$ do not depend on the incident illumination. Therefore, once we determine $a_\ell$ and $b_\ell$ using Eqs.~\eqref{eq:equation1}-\eqref{eq:equation3} for a specific illumination, such as a plane wave, we can subsequently explore the scattering features of the spherical object under general illumination conditions.

Interestingly, the dipolar Mie coefficients are bi-unequivocally determined by the electric and magnetic polarizabilities, oftently denoted as $\alpha_{\rm{E}}$ and $\alpha_{\rm{M}}$, respectively. For the sake of clarity, let us write the correspondence between Mie coefficients and polarizabilities~\cite{garcia2011strong}
\begin{align} \label{pol}
\alpha_{\rm{E}} = i \left(\frac{k^3}{6 \pi} \right)^{-1}a_1, &&  \alpha_{\rm{M}} = i \left(\frac{k^3}{6 \pi} \right)^{-1}b_1.
\end{align}
Equation~\eqref{pol} can be calculated from Eqs.~\eqref{eq:equation1}-\eqref{eq:equation3}, and thus,  our Stokes-polarimetry method can be employed to retrieve the electric and magnetic polarizabilities when dealing with dipolar Mie objects.


At this point, we show the accuracy of our method to solve the Maxwell equations in the radiation zone with a realistic material. Particularly, we consider a Gallium Phosphide (GaP) nanoparticle of radius $a = 75$ nm excited by a circularly polarized plane wave~\cite{shima2023gallium}. We select GaP as it is a material with high potential for metasurface-based 
devices operating across the visible, as it presents a high-refractive index ($\rm{m} > 3.3$) and negligible losses~\cite{https://doi.org/10.1002/lpor.202300553}.

Figure~\ref{GaP}a-d shows $\Re \{a_1 \}$, $\Im \{a_1 \}$, $\Re \{b_1 \}$, and $\Im \{b_1 \}$ calculated from a Stokes measurement at $\theta = 90^\circ$ and $\theta = 60^\circ$ and employing Mie theory. The scattering Mie coefficients calculated from the Stokes measurements at the specified angles exhibit excellent agreement with the exact calculations (and with each other) in the range  $475$ nm $< \lambda < 700$ nm.

It is worth noting that the results obtained from the Stokes vector measurement exhibit slight deviations from each other (and from the exact result) at shorter wavelengths, specifically in the range $450$ nm $< \lambda < 475$ nm. This deviation arises because, in this wavelength range, the scattering cannot be fully described by $\ell = m = 1$ due to the presence of the magnetic quadrupole.
Notably, our procedure robustly detects this quadrupole presence as the curves for $\theta = 90^\circ$ and $\theta = 60^\circ$ deviate in the range $450$ nm $< \lambda < 475$ nm. 
Our approach is reliable if the calculated coefficients remain identical regardless of the scattering angle $\theta$.  If the coefficients differ, then the scattering cannot be fully described by the selected values of $\ell$ and $m$ or/and the light-scattering system is not lossless.

Next, we show that the determination of the scattering Mie coefficients grants access to the internal Mie coefficients.

{\emph{Capturing the internal Mie coefficients.}}--- In 1908,  Gustav Mie solved the
scattering of a plane wave by a spherical object~\cite{mie1908beitrage}. Specifically, Mie determined the scattering $\{a_\ell, b_\ell \}$ and internal coefficients $\{d_\ell, c_\ell \}$.
The relation between $\{a_\ell, b_\ell \}$ and  $\{d_\ell, c_\ell \}$  and can be compactly written as~\cite{bohren2008absorption}
\begin{align}\label{int}
d_\ell = \frac{j_\ell(x) - a_\ell h^{(1)}_\ell(x)}{{\rm{m}} j_\ell({\rm{m}} x)}, && c_\ell = \frac{j_\ell(x) - b_\ell h^{(1)}_\ell(x)}{ j_\ell({\rm{m}} x)}.
\end{align}
Equation~\eqref{int} shows that the internal Mie coefficients $\{d_\ell, c_\ell \}$ can be determined from the scattering coefficients  $\{a_\ell, b_\ell \}$. Thus, the internal Mie coefficients can be captured using Eqs.~\eqref{eq:equation1}-\eqref{eq:equation3} particularized for spherical particles.

In Fig.~\ref{GaP}, we plot $\Re \{d_1\}$, $\Im \{d_1\}$, $\Re \{c_1\}$, and $\Im \{c_1\}$ using Eq.~\eqref{int}. For this calculation, we have employed $a_1$ and $b_1$, which, in turn, have been previously obtained using Eqs.~\eqref{eq:equation1}-\eqref{eq:equation3} at $\theta = 90^\circ$ and $\theta = 60^\circ$.
As could be expected, the calculation of the internal Mie coefficients shows an excellent agreement with the exact calculation in the wavelength interval $475 \; \rm{nm} < \lambda < 700 \;  \rm{nm}$.

As mentioned in the introduction, the determination of internal and scattering coefficients gives rise to the exact solution to Maxwell's equations. Since every electromagnetic physical magnitude originates from the electromagnetic field, we can also access, for instance, the exact internal dipolar moments, denoted as $\mathbf{p}$ and $\mathbf{m}$ in Ref.~\cite{alaee2018electromagnetic}.

In conclusion, we have presented a method that solves Maxwell's equations at all points in the radiation zone based on a single measurement of the Stokes parameters in the far-field. We have illustrated the accuracy of our method with one of the most studied systems in Nanophotonics: a spherical nanoparticle excited by a plane wave. 
In this setting, we have also determined the internal Mie coefficients, solving Maxwell's equations at all points in the space.


To the best of our knowledge, this study represents the first method capable of solving Maxwell's equations experimentally and from a measurement of the Stokes parameters. This feature endows the Stokes parameters, mostly used to get insight into the polarization state of the electromagnetic radiation, an even more fundamental role in the electromagnetic scattering theory. 
Consequently, our findings, supported by analytical theory and exact numerical simulations, can find applications in all branches of Nanophotonics and Optics.

\clearpage

\appendix 

\section{The Stokes method}

The scattered  field $\E (k \r)$ in the radiation zone (when $ kr \rightarrow \infty$) can be expressed as~\cite{jackson1999electrodynamics}  
\begin{equation} \label{E_far_1}
\lim_{kr \rightarrow \infty} \E (k \r) = \left[ E_\theta \hat{\mathbf{e}}_{{{\theta}}} + E_\varphi \hat{\mathbf{e}}_{{{\varphi}}}\right], 
\end{equation}
where 
\begin{equation} \label{key_1}
E_\theta = {E_0} \sum_{\ell m} \bar{C}_{\ell  m} (kr, \varphi) \left[  a_{\ell m} \tau_{\ell m} ({\theta}) - im b_{\ell m} \pi_{\ell m} ({\theta}) \right],
\end{equation}
\begin{equation}\label{key_2}
E_\varphi = {E_0} \sum_{\ell m}  \bar{C}_{\ell  m} (kr, \varphi) \left[ im  a_{\ell m} \pi_{\ell m} ({\theta}) + b_{\ell m} \tau_{\ell m} ({\theta}) \right].
\end{equation}
Here $a_{\ell m}$ and $b_{\ell m}$ denote the (dimensionless) electric and magnetic scattering coefficients, respectively, $\ell$ and $m$ being the multipolar order and total angular momentum, respectively. Additionally, 
$E_0$ is the amplitude of the incident wavefield,  $k$ is the radiation wavenumber,  $r = |\r|$ denotes the observation distance to the center of the object, and  $\theta$ and $\varphi$ denote the scattering and azimuthal angles, respectively. Moreover, we have defined~\footnote{Interestingly, $\pi_{lm}(\theta)$ and $\tau_{l m} (\theta)$  are real-valued functions that Bohren and Huffman defined to tackle the absorption and scattering by a sphere for $m = 1$  (see Eq. 4.46 of Ref.~\cite{bohren2008absorption}).} 
\begin{align} \label{chi_tau}
\pi_{\ell m}(\theta) =  \frac{P^{m}_{\ell }(\cos \theta)}{\sin \theta}, &&    \tau_{\ell m}(\theta)  = \frac{d P^{m}_{\ell }(\cos \theta)}{d{\theta}},
\end{align}
where $P^{m}_{\ell}(\cos \theta)$ are the Associated Legendre Polynomials~\cite{jackson1999electrodynamics} and
\begin{equation} \label{constant}
\bar{C}_{\ell  m} (kr, \varphi)= \frac{e^{ikr}}{kr} \left[ \frac{(-i)^{\ell + 2}}{\sqrt{\ell ( \ell + 1) }} \sqrt{\frac{2 \ell +1}{4\pi}\frac{(\ell - m)!}{(\ell+m)!}} \right]e^{im \varphi}.
\end{equation}

At this stage, we have introduced all the requirements to calculate the Stokes vector $\mathbf{S} = \{s_0, s_1, s_2, s_3 \}$. 
Now, by considering that the object can be described by a single multipolar order $\ell$ and total angular momentum $m$, it can be shown that~\cite{olmos2023stokes}
\begin{align} \label{s0_new}
\tilde{s}_0 &=  (|a_{\ell m}|^2 + |b_{\ell m}|^2)\gamma_{\ell m}(\theta)  - 4 \Im \{a_{\ell m} b^*_{\ell m} \} \eta_{\ell m}(\theta), \\
\tilde{s}_1 &= (|a_{\ell m}|^2 - |b_{\ell m}|^2)\nu_{\ell m}(\theta), \\ \label{mystery}
\tilde{s}_2 &= -2 \Re \{a_{\ell m} b^*_{\ell m} \} \nu_{\ell m}(\theta), \\ \label{s3_new}
\tilde{s}_3 &= 2 \left[\Im \{a_{\ell m} b^*_{\ell m} \}\gamma_{\ell m}(\theta) -(|a_{\ell m}|^2 + |b_{\ell m}|^2)\eta_{\ell m}(\theta) \right].
\end{align}
Here, we have defined  ${\mathbf{S}} = |E_0|^2|\bar{C}_{\ell m}(kr, \varphi)|^2 \mathbf{\tilde{S}}$ along with
\begin{align} \label{def_1}
 \gamma_{\ell m}(\theta) &= \left[ \tau_{\ell m} ^2(\theta) + m^2 \pi_{\ell m} ^2(\theta)\right], \\
 \eta_{\ell m}(\theta)  &= m   \tau_{\ell m} (\theta)\pi_{\ell m} (\theta), \\ \label{def_2}
 \nu_{\ell m}(\theta) &= \left[ \tau_{\ell m} ^2(\theta) - m^2 \pi_{\ell m} ^2(\theta)\right]. 
\end{align}
Now, we can rewrite 
Eqs.~\eqref{s0_new}-\eqref{s3_new} as
\begin{equation} \label{compact_appendix}
\mathbf{D}_{\ell m} =   U_{\ell m}  {{\mathbf{S}}},
\end{equation}
\begin{align} \label{matrix_change}
U_{\ell m} = \frac{1}{A_{\ell m}}   
    \begin{pmatrix}
    \gamma_{\ell m} & \nu_{\ell m}& 0 & 2\eta_{\ell m}\\
    \gamma_{\ell m}& -\nu_{\ell m}& 0 & 2\eta_{\ell m} \\
       0 & 0 & -\nu_{\ell m} & 0\\
    2\eta_{\ell m} & 0 & 0 &  \gamma_{\ell m}
    \end{pmatrix},
\end{align}
with $A_{\ell m} = 2|E_0|^2|\bar{C}_{\ell m}|^2 \nu^2_{\ell m}$, and 
\begin{align} \label{vector}
\mathbf{D}_{\ell m} = 
 \begin{pmatrix}
    |a_{\ell m}|^2\\
    |b_{\ell m}|^2 \\ \Re \{a_{\ell m} b^*_{\ell m} \}\\
    \Im \{a_{\ell m} b^*_{\ell m} \}
    \end{pmatrix}, &&
    {\mathbf{{S}}} = 
 \begin{pmatrix}
    {s}_0\\
    {s}_1\\ {s}_2\\
    {s}_3
    \end{pmatrix}.
\end{align}

\section{The beam shape coefficients of a plane wave} \label{A_1}

The beam-shape coefficients of a plane wave propagating in the $z$-direction, $\E_{\rm{inc}} = E_0 e^{ikz} \left(p_x, p_y, 0 \right)$, are well-known and given by~\cite{jackson1999electrodynamics} 
\begin{align} \label{pw}
g^{\rm{e}}_{\ell, \pm 1} = G_\ell \left(\mp ip_x - p_y \right), && g^{\rm{m}}_{\ell, \pm 1} = G_\ell \left(p_x \mp i p_y \right),     
\end{align}
where $G_\ell = i^\ell \sqrt{ \pi (2 \ell +1)}$. The polarization is described by the components of the Jones vector $p_x$ and $p_y$, which satisfies $|p_x|^2 + |p_y|^2 = 1$. For the sake of simplicity, let us fix the incident wavefield to a circularly polarized plane. A circularly polarized plane wave satisfies $m = p$, where $p = \pm 1$ is the helicity (handedness) of the wave. For instance, a left-handed polarized plane wave carries $p=+1$ and thus $m = +1$. The components of the Jones vector for a left-handed polarized plane wave are given by $p_x = -i p_y = 1/ \sqrt{2}$. Hence, the beam shape coefficients of a left-handed circularly polarized plane wave are given by
\begin{align} \label{pw_left}
g^{\rm{e}}_{\ell 1} = -i \bar{G}_\ell , && g^{\rm{m}}_{\ell 1} = \bar{G}_\ell 
\end{align}
where $\bar{G}_\ell =  \sqrt{2} {G}_\ell$. These are the beam-shape coefficients used in the main manuscript.

\section{A lossless magnetodielectric spherical nanoparticle} \label{lossless}
The T-matrix of a spherical particle is diagonal and satisfies $a_{\ell m} = -g^{\rm{e}}_{\ell m} a_\ell$ and $b_{\ell m} = -g^{\rm{m}}_{\ell m} b_\ell$, $a_\ell$ and $b_\ell$ being the electric and magnetic Mie coefficients, respectively~\cite{mie1908beitrage}. Taking these relations into account, we can write from the left side of Eq.~12
\begin{align} \label{amp_e}
|a_{\ell m}|^2 &= |g^{\rm{e}}_{\ell m}|^2  |a_{\ell} |^2, \\
|b_{\ell m}|^2 &= |g^{\rm{m}}_{\ell m}|^2  |b_{\ell} |^2, \\
\Re \{ a_{\ell m} b^*_{\ell m} \} &= \Re \{ a_{\ell} b^*_{\ell} \} \Re \{ g^{\rm{e}}_{\ell m}{g^{\rm{m}}_{\ell m}}^* \} - \Im \{ a_{\ell} b^*_{\ell} \} \Im \{ g^{\rm{e}}_{\ell m}{g^{\rm{m}}_{\ell m}}^* \}, \\ \label{imag_em}
\Im \{ a_{\ell m} b^*_{\ell m} \} &= \Re \{ a_{\ell} b^*_{\ell} \} \Im \{ g^{\rm{e}}_{\ell m}{g^{\rm{m}}_{\ell m}}^* \} + \Im \{ a_{\ell} b^*_{\ell} \} \Re \{ g^{\rm{e}}_{\ell m}{g^{\rm{m}}_{\ell m}}^* \}.
\end{align}
Now, the electric and magnetic Mie coefficients of a lossless spherical particle can be written in the scattering phase-shift notation~\cite{hulst1957light}. That is,
\begin{align} \label{phase_shifts}
a_{\ell} = i \sin \alpha_\ell e^{-i \alpha_\ell}, &&    b_{\ell} = i \sin \beta_\ell e^{-i \beta_\ell}, 
\end{align}
where $\alpha_\ell$ and $\beta_\ell$ are real in the absence of losses. In this setting, it can be shown that $\Re \{ a_{\ell} \} = |a_{\ell}|^2 = \sin^2 \alpha_\ell$ and $\Re \{ b_{\ell} \} = |b_{\ell}|^2 = \sin^2 \beta_\ell$. Note that $|a_\ell|>0$ and $ |b_\ell|>0$ implies that $0<\alpha_\ell<\pi$ and $0<\beta_\ell<\pi$, respectively.

Inserting Eq.~\eqref{phase_shifts} into Eqs.~\eqref{amp_e}-\eqref{imag_em} yield
\begin{align} \label{amp_e1}
|a_{\ell m}|^2 &= |g^{\rm{e}}_{\ell m}|^2  \sin^2 \alpha_\ell, \\
|b_{\ell m}|^2 &= |g^{\rm{m}}_{\ell m}|^2  \sin^2 \beta_\ell, \\
\Re \{ a_{\ell m} b^*_{\ell m} \} &= \sin \alpha_\ell \sin \beta_\ell \cos(\alpha_\ell - \beta_\ell) \Re \{ g^{\rm{e}}_{\ell m}{g^{\rm{m}}_{\ell m}}^* \} \\ \nonumber
&+ \sin \alpha_\ell \sin \beta_\ell \sin(\alpha_\ell - \beta_\ell) \Im \{ g^{\rm{e}}_{\ell m}{g^{\rm{m}}_{\ell m}}^* \}, \\ \label{imag_em1}
\Im \{ a_{\ell m} b^*_{\ell m} \} &= \sin \alpha_\ell \sin \beta_\ell \cos(\alpha_\ell - \beta_\ell) \Im \{ g^{\rm{e}}_{\ell m}{g^{\rm{m}}_{\ell m}}^* \} \\ \nonumber
&- \sin \alpha_\ell \sin \beta_\ell \sin(\alpha_\ell - \beta_\ell) \Re \{ g^{\rm{e}}_{\ell m}{g^{\rm{m}}_{\ell m}}^* \}.
\end{align}
These equations are valid for a lossless spherical particle. For the sake of simplicity, let us insert the beam shape coefficients of a left-handed circularly polarized plane wave (see Eq.~\eqref{pw_left}) into Eqs.~\eqref{amp_e1}-\eqref{imag_em1}. Moreover, let us assume $\ell = 1$ (dipolar response). Taking into account this setting, we arrive to
\begin{align} \label{amp_e2}
|a_{1 1}|^2 &= |\bar{G}_1|^2  \sin^2 \alpha_1, \\
|b_{1 1}|^2 &= |\bar{G}_1|^2  \sin^2 \beta_1, \\
\Re \{ a_{1 1} b^*_{1 1} \} &=  - |\bar{G}_1|^2\sin \alpha_1 \sin \beta_1, \sin(\alpha_1 - \beta_1),  \\ \label{imag_em2}
\Im \{ a_{1 1} b^*_{11} \} &=  - |\bar{G}_1|^2\sin \alpha_1 \sin \beta_1 \cos(\alpha_1 - \beta_1). 
\end{align}
These equations can be further simplified, yielding
\begin{align} \label{amp_e3}
|a_{1 1}| &= |\bar{G}_1|  \sin \alpha_1, \\
|b_{1 1}| &= |\bar{G}_1|  \sin \beta_1, \\ \label{amp_em3}
\text{atan2} \left[\frac{\Re \{ a_{1 1} b^*_{1 1} \}}{\Im \{ a_{1 1} b^*_{1 1} \}} \right] &= \alpha_1 - \beta_1 .
\end{align}
These equations can be solved from a single measurement of the Stokes parameters. As we explain in the main text, this Stokes measurement yields $a_{11}$, $b_{11}$, $\Re \{a_{11} b^*_{11} \}$, $\Im \{a_{11} b^*_{11} \}$. Then, by using Eq.~\eqref{amp_e3}-\eqref{amp_em3} we can unambiguously obtain
$\alpha_1$ and $\beta_1$. Then, by inserting these values into Eq.~\eqref{phase_shifts} we can obtain the electric and magnetic Mie coefficients for $\ell =1$.  This is the procedure that we have followed to calculate Figs 1-2.

\clearpage


\section*{References}
\bibliography{Bib_tesis}

\clearpage

\end{document}